\documentclass[a4paper,11pt]{article}
\usepackage{jinstpub} 
\usepackage{lineno}
\usepackage{lineno,hyperref}
\usepackage{color}
\usepackage{gensymb}
\usepackage{CJKutf8}
\usepackage{multirow}
\usepackage{siunitx}
\usepackage{makecell}
\usepackage{graphicx}
\usepackage{footnote}
\usepackage{subcaption}

\title{\boldmath A Novel Diamond-like Carbon based photocathode for PICOSEC Micromegas detectors}

\author[a,b]{X. Wang,}
\author[c]{R. Aleksan,}
\author[d]{Y. Angelis,}
\author[e]{J. Bortfeldt,}
\author[f]{F. Brunbauer,}
\author[g,h]{M. Brunoldi,}
\author[d]{E. Chatzianagnostou,}
\author[i]{J. Datta,}
\author[j]{K. Degmelt,}
\author[k]{G. Fanourakis,}
\author[g,h,1]{D. Fiorina\note{Now at Gran Sasso Science Institute, Viale F. Crispi, 7 67100 L’Aquila, Italy.},}
\author[f,l]{K.J. Floethner,}
\author[m]{M. Gallinaro,}
\author[n]{F. Garcia,}
\author[c]{I. Giomataris,}
\author[j]{K. Gnanvo,}
\author[c,2]{F.J. Iguaz\note{Now at SOLEIL Synchrotron, L’Orme des Merisiers, Départementale 128, 91190 SaintAubin, France.},}
\author[f]{D. Janssens,}
\author[c]{A. Kallitsopoulou,}
\author[o]{M. Kovacic,}
\author[j]{B. Kross,}
\author[c]{P. Legou,}
\author[f,p]{M. Lisowska,}
\author[a,b]{J. Liu,}
\author[f,c,3]{I. Maniatis\note{Now at Department of Particle Physics and Astronomy, Weizmann Institute of Science, Rehovot, 7610001, Israel.},}
\author[j]{J. McKisson,}
\author[a,b]{Y. Meng,}
\author[f,q]{H. Muller,}
\author[f]{E. Oliveri,}
\author[f,r]{G. Orlandini,}
\author[j]{A. Pandey,}
\author[c]{T. Papaevangelou,}
\author[s]{M. Pomorski,}
\author[f]{L. Ropelewski,}
\author[d,t]{D. Sampsonidis,}
\author[f]{L. Scharenberg,}
\author[f]{T. Schneider,}
\author[u]{L. Shang,}
\author[a,b]{M. Shao,}
\author[c,4]{L. Sohl\note{Now at TÜV NORD EnSys GmbH Co. KG.},}
\author[f]{M. van Stenis,}
\author[v]{Y. Tsipolitis,}
\author[d,t]{S.E. Tzamarias,}
\author[w]{A. Utrobicic,}
\author[g,h]{I. Vai,}
\author[f]{R. Veenhof,}
\author[g,h]{P. Vitulo,}
\author[f,x]{S. White,}
\author[j]{W. Xi,}
\author[a,b]{Z. Zhang,}
\author[a,b,5]{and Y. Zhou\note{Corresponding author.}}

\affiliation[a]{State Key Laboratory of Particle Detection and Electronics, University of Science and Technology of China, Hefei 230026, China}
\affiliation[b]{Department of Modern Physics, University of Science and Technology of China, Hefei 230026, China}
\affiliation[c]{IRFU, CEA, Université Paris-Saclay, F-91191 Gif-sur-Yvette, France}
\affiliation[d]{Department of Physics, Aristotle University of Thessaloniki, University Campus, GR-54124, Thessaloniki, Greece}
\affiliation[e]{Department for Medical Physics, Ludwig Maximilian University of Munich, Am Coulombwall 1, 85748 Garching, Germany}
\affiliation[f]{European Organisation for Nuclear Research (CERN), CH-1211, Geneve 23, Switzerland}
\affiliation[g]{Dipartimento di Fisica, Università di Pavia, Via Bassi 6, 27100 Pavia (IT)}
\affiliation[h]{INFN Sezione di Pavia, Via Bassi 6, 27100 Pavia (IT)}
\affiliation[i]{Stony Brook University, Dept. of Physics and Astronomy, Stony Brook, NY 11794-3800, USA}
\affiliation[j]{Jefferson Lab, 12000 Jefferson Avenue, Newport News, VA 23606, USA}
\affiliation[k]{Institute of Nuclear and Particle Physics, NCSR Demokritos, GR-15341 Agia Paraskevi, Attiki, Greece}
\affiliation[l]{Helmholtz-Institut für Strahlen- und Kernphysik, University of Bonn, Nußallee 14–16, 53115 Bonn, Germany}
\affiliation[m]{Laboratório de Instrumentacão e Física Experimental de Partículas, Lisbon, Portugal}
\affiliation[n]{Helsinki Institute of Physics, University of Helsinki, FI-00014 Helsinki, Finland}
\affiliation[o]{Faculty of Electrical Engineering and Computing, University of Zagreb, 10000 Zagreb, Croatia}
\affiliation[p]{Université Paris-Saclay, F-91191 Gif-sur-Yvette, France}
\affiliation[q]{Physikalisches Institut, University of Bonn, Nußallee 12, 53115 Bonn, Germany}
\affiliation[r]{Friedrich-Alexander-Universität Erlangen-Nürnberg, Schloßplatz 4, 91054 Erlangen, Germany}
\affiliation[s]{CEA-LIST, Diamond Sensors Laboratory, CEA Saclay, F-91191 Gif-sur-Yvette, France}
\affiliation[t]{Center for Interdisciplinary Research and Innovation (CIRI-AUTH), Thessaloniki 57001, Greece}
\affiliation[u]{State Key Laboratory of Solid Lubrication, Lanzhou Institute of Chemical Physics, Chinese Academy of Science, Lanzhou 730000, China}
\affiliation[v]{National Technical University of Athens, Athens, Greece}
\affiliation[w]{Ruđer Bošković Institute, Bijenička cesta 54, 10000 Zagreb, Croatia}
\affiliation[x]{University of Virginia, Charlottesville, VA, 22904, United States of America}

\emailAdd{zhouyi@mail.ustc.edu.cn}
\abstract{The PICOSEC Micromegas (MM) detector is a precise timing gaseous detector based on a MM detector operating in a two-stage amplification mode and a Cherenkov radiator. Prototypes equipped with cesium iodide (CsI) photocathodes have shown promising time resolutions as precise as 24 picoseconds (\si{ps}) for Minimum Ionizing Particles. However, due to the high hygroscopicity and susceptibility to ion bombardment of the CsI photocathodes, alternative photocathode materials are needed to improve the robustness of PICOSEC MM. Diamond-like Carbon (DLC) film have been introduced as a novel robust photocathode material, which have shown promising results. A batch of DLC photocathodes with different thicknesses were produced and evaluated using ultraviolet light. The quantum efficiency measurements indicate that the optimized thickness of the DLC photocathode is approximately 3 \si{nm}. Furthermore, DLC photocathodes show good resistance to ion bombardment in aging test compared to the CsI photocathode. Finally, a PICOSEC MM prototype equipped with DLC photocathodes was tested in muon beams. A time resolution of around 42 \si{ps} with a detection efficiency of 97\% for 150 \si{GeV/c} muons were obtained. These results indicate the great potential of DLC as a photocathode for the PICOSEC MM detector. }


\begin{document}
\maketitle
\flushbottom

\section{Introduction}
The development of a new generation of particle detectors with precise timing performance has been driven by the challenging environments of future High Energy Physics experiments, including severe pile-up effects in the upgraded High Luminosity Large Hadron Collider. Time resolution on the order of tens of picoseconds (\si{ps}), as well as high granularity, long-term stability, robustness and large area coverage\cite{white2013experimental,colaleo20212021}, are required. The study of detectors based on micro-pattern gaseous detectors (MPGDs) provide potential solutions since the advantages of MPGDs, including high-rate capability, radiation resistance and the ability to cover large area at low cost. MPGDs with precise timing performance have a wide range of applications, such as time-of-flight measurement, particle identification and suppression of pile-up effects in high luminosity environments\cite{chekanov2020physics,va2017pid}.\par

\begin{figure}[ht]
  \centering
  \begin{subfigure}[b]{0.49\textwidth}
    \includegraphics[width=1.\textwidth]{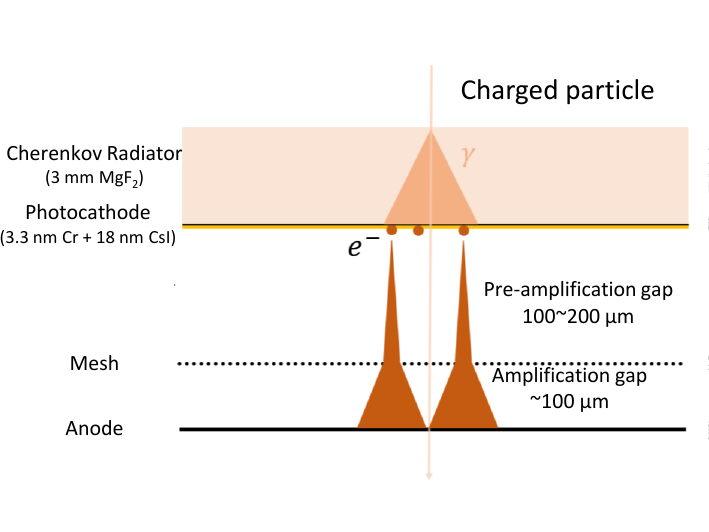}
    \caption{}
    \label{figure:sub1}
  \end{subfigure}
  \hfill 
  \begin{subfigure}[b]{0.49\textwidth}
    \includegraphics[width=0.9\textwidth]{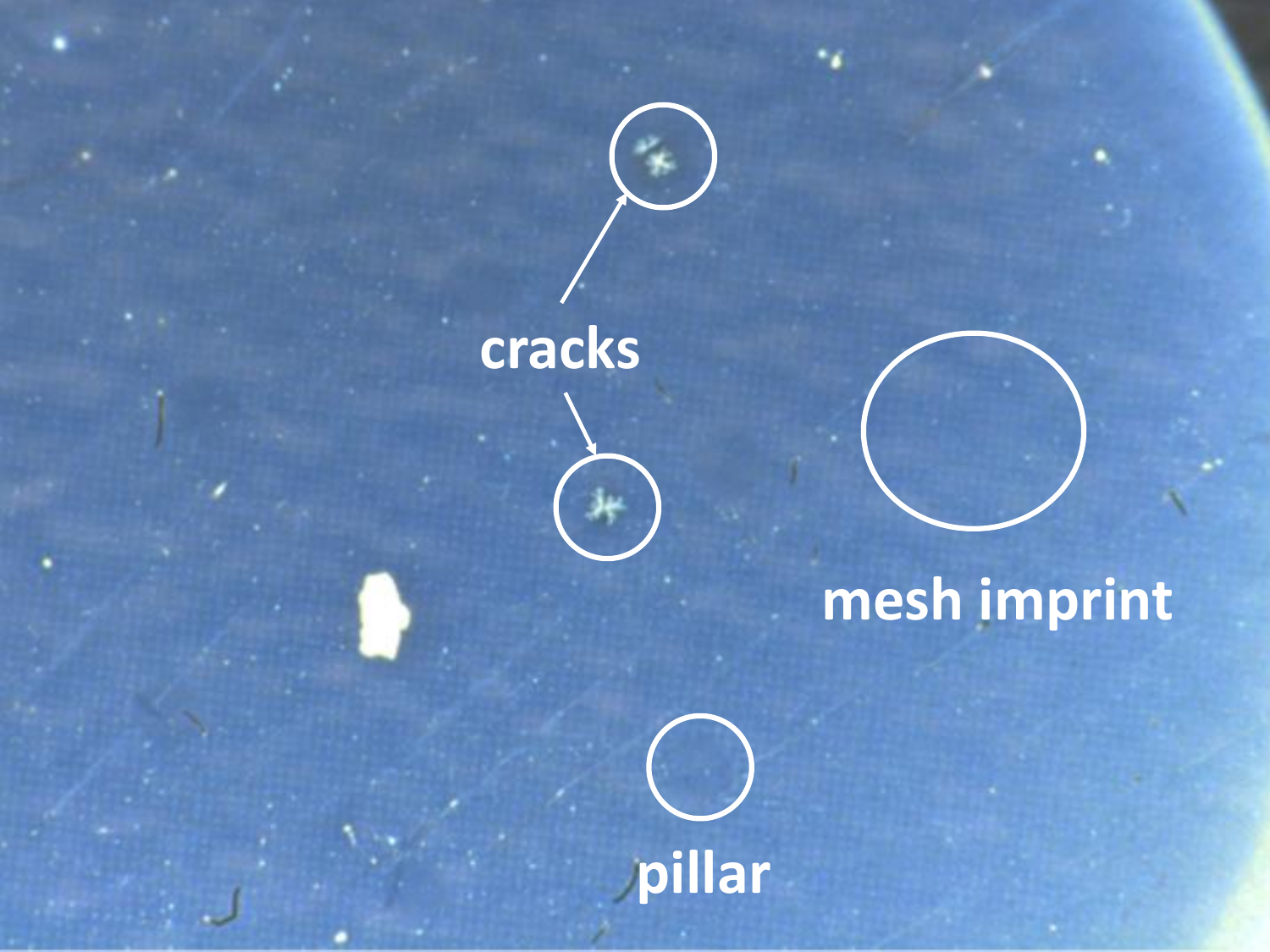}
    \caption{}
    \label{figure:sub2}
  \end{subfigure}
  \caption{Schematic view of PICOSEC MM detector concept (a). The microscopic image of CsI photocathode after beam test (b)}
  \label{figure:test}
\end{figure}

A concept based on a two-stage Micromegas (MM) structure and Cherenkov radiation detection, called PICOSEC Micromegas (PICOSEC MM), was proposed in 2015\cite{papaevangelou2018fast}. Figure~\ref{figure:sub1} illustrates the detector concept. A Cherenkov radiator (3 \si{mm} MgF$_2$) coated with a photocathode, consisting of 18 \si{nm} Cesium Iodide (CsI) and 3.3 \si{nm} metallic base, is coupled to the MM. Ultraviolet (UV) photons are generated when a charged particle passes through the radiator, and then the photoelectrons (PEs) are simultaneously extracted from the photocathode. Due to the high electric field applied between the photocathode and the mesh, the PEs are first multiplied in the pre-amplification gap. Then, a portion of the avalanche electrons can pass through the mesh and are further amplified in the amplification gap. The working gas used by this detector is a gas mixture of 80\% Ne, 10\% C$_2$H$_6$, 10\% CF$_4$. A lot of studies, including modeling and fabrication of prototypes, have been conducted within the PICOSEC MM collaboration\cite{papaevangelou2018fast,bortfeldt2018picosec,bortfeldt2021modeling}. For the prototype with CsI photocathodes, a time resolution of 24 \si{ps} at a mean yield of 10.4 PEs has been obtained for 150 \si{GeV/c} muon beams at the CERN SPS H4 secondary line\cite{bortfeldt2018picosec}. However, due to the high electric field in the pre-amplification gap, substantial feedback ions bombard the CsI photocathode, causing an aging problem\cite{xuwang1}. Figure~\ref{figure:sub2} shows the microscopic image of the CsI photocathode after a beam test. It is noticeable that there is a mesh projection due to ion impacts, along with the presence of circular area attributed to the pillars of MM. The performance degeneration in another aspect is the cracks on the CsI layer caused by the sparks occurring within the gap. Additionally, a dry environment is required for storage and operation due to its hygroscopic and deliquescent nature\cite{xie2012influence,razin1998influence}.\par

\begin{figure}[ht]
\centering
\includegraphics[width=.5\textwidth]{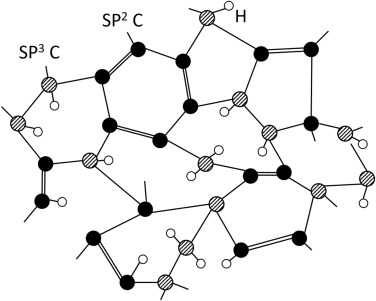}
\caption{The atom structure of DLC, which contains both sp$^3$ (diamond-structure) and sp$^2$ (graphite-structure)\cite{zhou2019fabrication}}
\label{DLCstructure}
\end{figure}

To ensure the long-term operation of the PICOSEC MM, alternative photocathodes with chemical stability, resistance to ion bombardment, and suitable quantum efficiency (QE) are needed. Diamond films can be employed for the detection of UV photons with reasonable sensitivity, without significant problems of aging and radiation damage\cite{nitti2008performance}. However, the production of diamond films has several limitations such as high temperature requirements and small area. Diamond-like Carbon (DLC) is a kind of novel resistive material with metastable amorphous structures containing both diamond-structure and graphite-structure carbon atoms\cite{grill1999diamond}, as shown in figure~\ref{DLCstructure}. It has excellent electrical properties, suitable sensitivity, chemical stability and thermal stability. DLC films have been studied for the photocathode applications such as electron guns\cite{balalykin2014detailed,paraliev2010experimental}. Based on this, we proposed DLC as the photocathode material for PICOSEC MM. In this paper, we present the study of the DLC photocathode.  The magnetron sputtering technology used to fabricate the DLC photocathode is described in section~\ref{Production of DLC photocathode}. Subsequently, the UV light was used to characterise the QE performance and aging effect of DLC phtotcathodes, as shwon in section~\ref{QE measurement} and section~\ref{Aging test}. Then, section~\ref{Beam test} shows the beam tests of a PICOSEC MM prototype equipped with DLC photocathodes. The last section is a summary drawn from this paper. \par

\section{Production of DLC photocathode}
\label{Production of DLC photocathode}
\begin{figure}[ht]
\centering
\includegraphics[width=.6\textwidth]{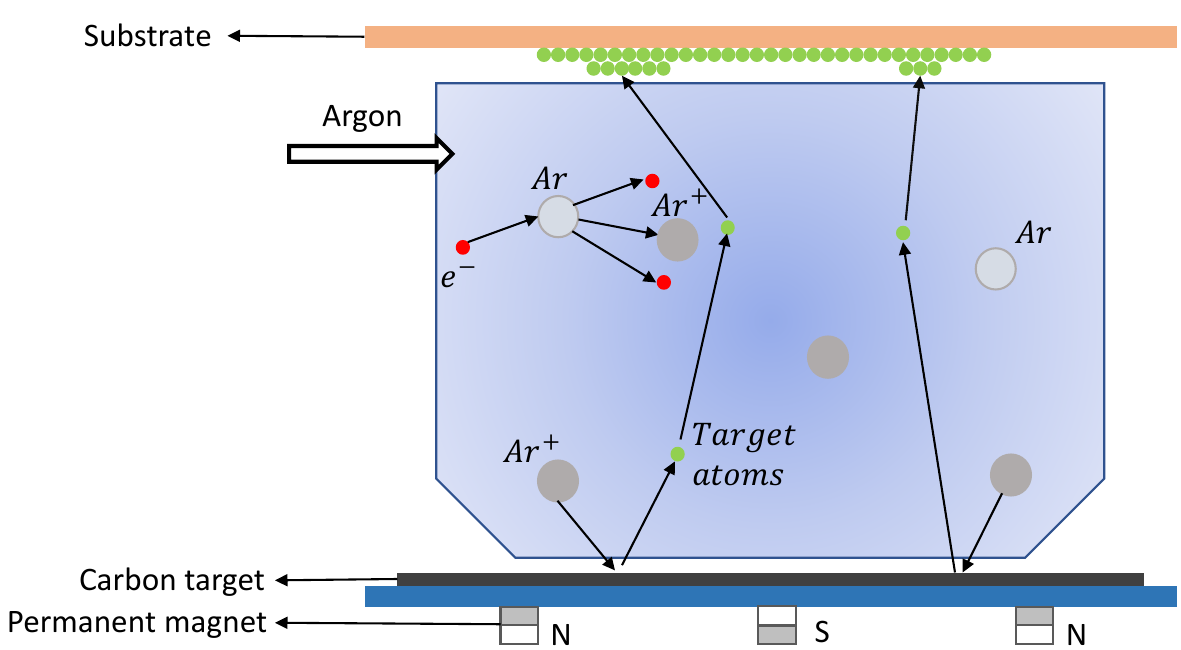}
\caption{Schematic diagram of DLC deposition by the magnetron sputtering technology}
\label{DLCcoating}
\end{figure}

The magnetron sputtering technology\cite{kelly2000magnetron} is an effective method for DLC deposition at low temperature. It provides a deposition process with good reproducibility and controllability, ensuring accuracy for the thickness and uniformity of the DLC film. The DLC film is directly deposited onto the surface of a MgF$_2$ crystal to form the DLC photocathode. A schematic diagram of the deposition process is illustrated in figure~\ref{DLCcoating}. Several permanent magnets are arranged on the back side of a graphite target to produce closed magnetic field lines around the surface of the target. Before commencing the deposition process, the sputtering chamber is evacuated to the preset vacuum level, afterward argon is filled into the chamber. As the bias voltage between the substrate and the target ramps up, glow discharges occur and produce primary electrons. These primary electrons are accelerated by the electric field and ionize the argon molecules into ions and secondary electrons. The argon ions are accelerated in the field, then bombard the target and sputter out a number of carbon atoms or clusters. These sputtered carbon atoms or clusters are then deposited onto the surface of the substrate, forming a thin film. The secondary electrons are bounded inside the plasma region near the target surface by the Lorentz force of the magnetic field. Continuously ionizing the argon molecules, they producing argon ions that further bombard the target.\par

\begin{figure}[ht]
  \centering
  \begin{subfigure}[b]{0.49\textwidth}
    \includegraphics[width=0.9\textwidth]{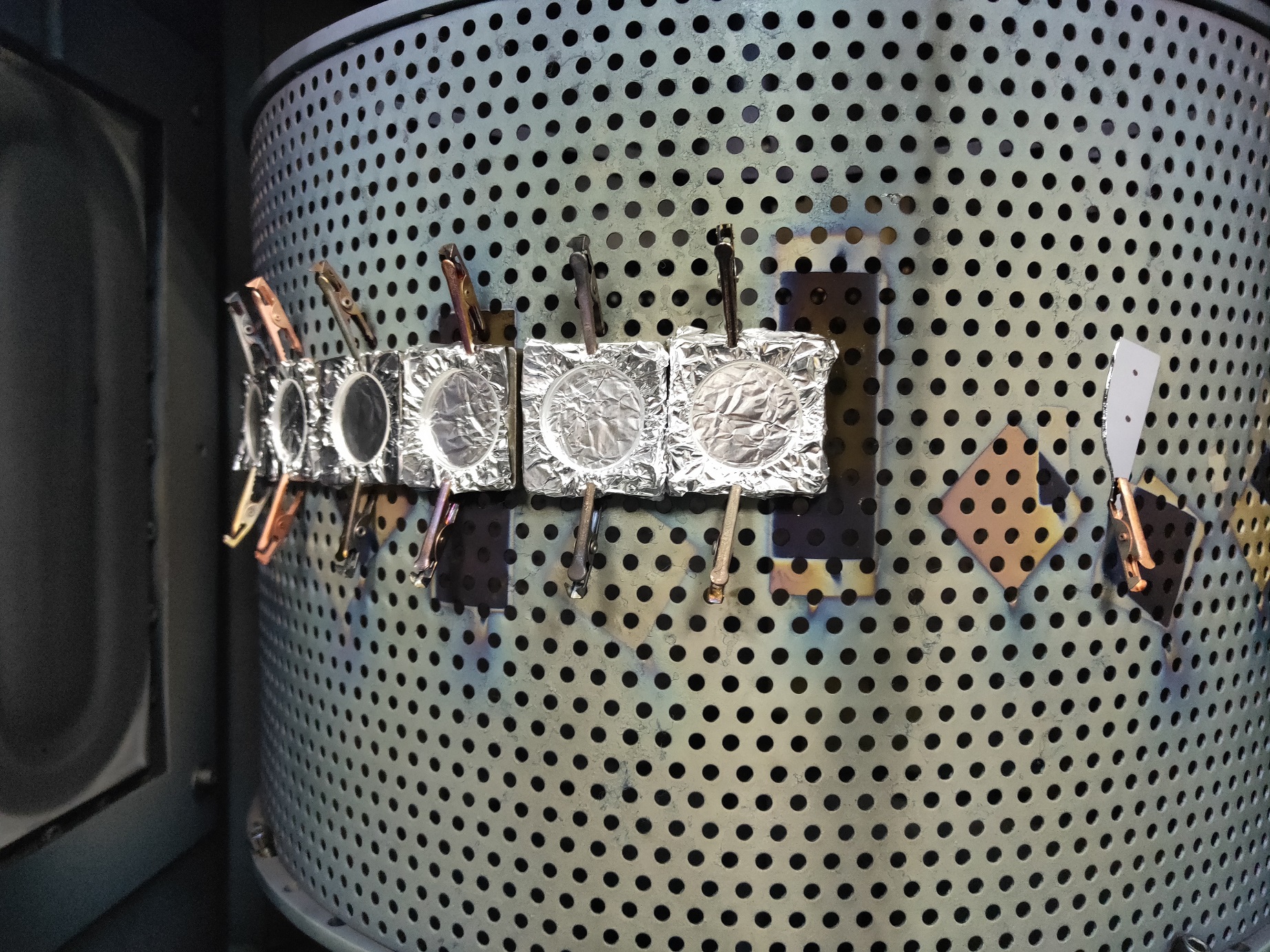}
    \caption{}
    \label{DLCsamples}
  \end{subfigure}
  \hfill 
  \begin{subfigure}[b]{0.49\textwidth}
    \includegraphics[width=0.9\textwidth]{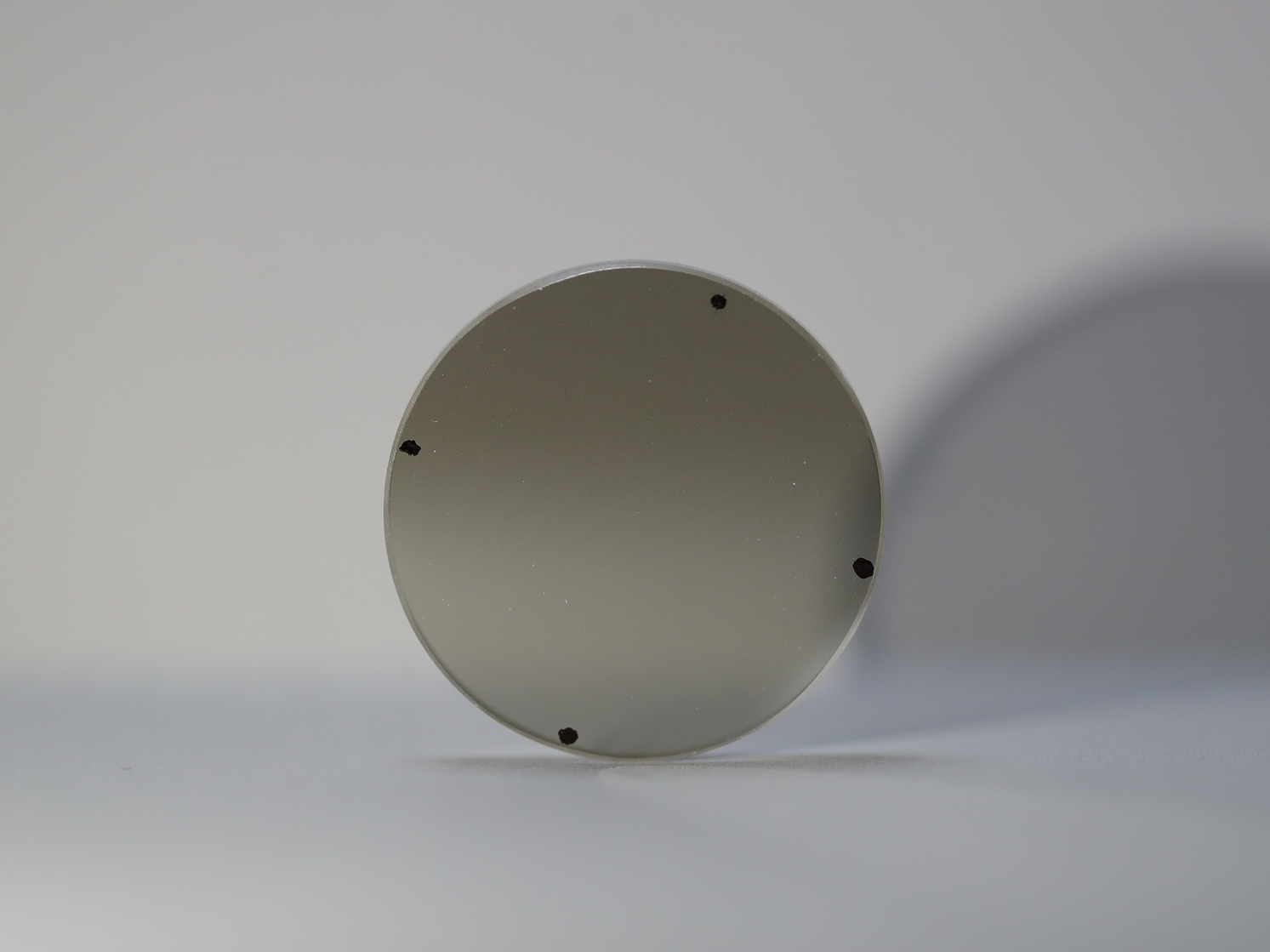}
    \caption{}
    \label{DLCphotocathode}
  \end{subfigure}
  \caption{MgF$_2$ crystals are fixed in the coating chamber (a) and a sample of DLC photocathode (b)}
  \label{DLCcoatingsamples}
\end{figure}

The coating machine we used is Teer650 (Teer Coating Ltd) at Lanzhou Institute of Chemical Physics. During the coating process, the MgF$_2$ crystal is fixed in the aluminum holders before being positioned within the chamber, as shown in figure~\ref{DLCsamples}. The coating vacuum is maintained at a level better than $1\times10^{-6}$ \si{Torr}. A series of DLC photocathodes with the film thicknesses varying from 1 \si{nm} to 10 \si{nm} were produced by adjusting the deposition time while keeping other conditions constant. As shown in figure~\ref{DLCphotocathode}, the DLC film, which is marked with four black dots, is deposited on a 3 \si{mm} thick MgF$_2$ crystal.\par

\section{QE measurement}
\label{QE measurement}
A vacuum ultraviolet (VUV) QE measurement setup (ASSET) was developed by the Gaseous Detectors Development Group (GDD) at CERN. The QE measurement in the transmissive model of this setup was used to characterise DLC photocathodes. Various photocathodes with different thicknesses were measured to obtain the optimized parameters.\par

\subsection{Setup}
In ASSET setup, a VUV monochromator (McPherson 234/302) is used to select the wavelength of the UV photons produced by a deuterium lamp with an accuracy of 0.1 \si{nm}. The UV photons are split into two beams by a beam splitter. One of them irradiates on a photomultiplier tube (ET-enterprises 9403B) (PMT$_1$), while the other one enters a measurement chamber and directly irradiates on another PMT (PMT$_2$) or passes through the sample to be measured. An overview of the ASSET setup is shown in \cite{Marta}. \par
\begin{figure}[ht]
  \centering
  \begin{subfigure}[b]{0.49\textwidth}
    \includegraphics[width=\textwidth]{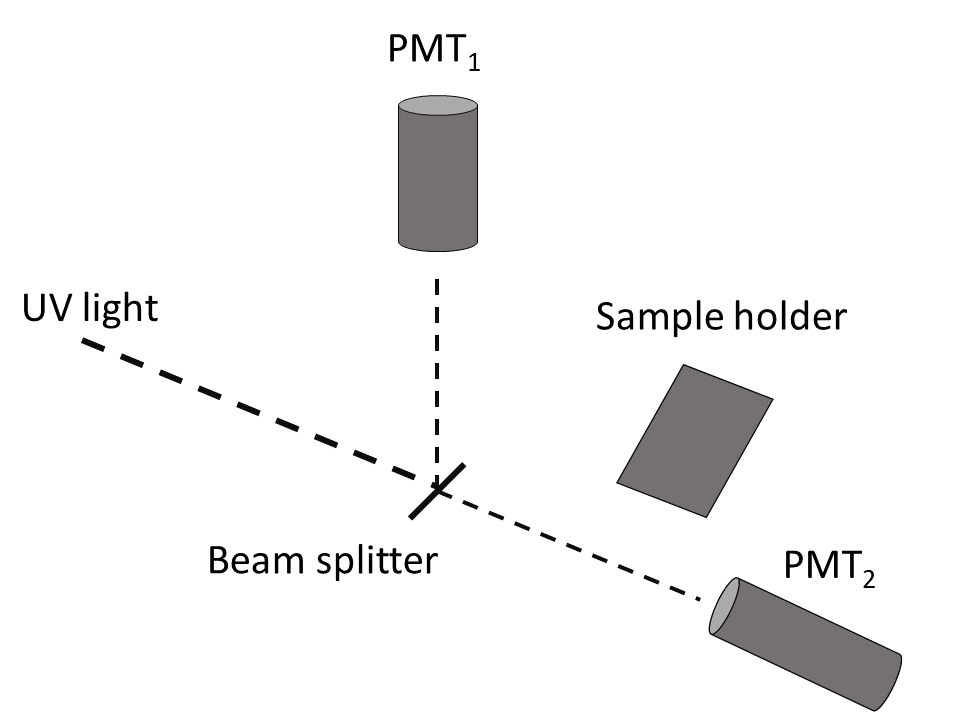}
    \caption{}
    \label{ASSET_step1}
  \end{subfigure}
  \hfill
  \begin{subfigure}[b]{0.49\textwidth}
    \includegraphics[width=\textwidth]{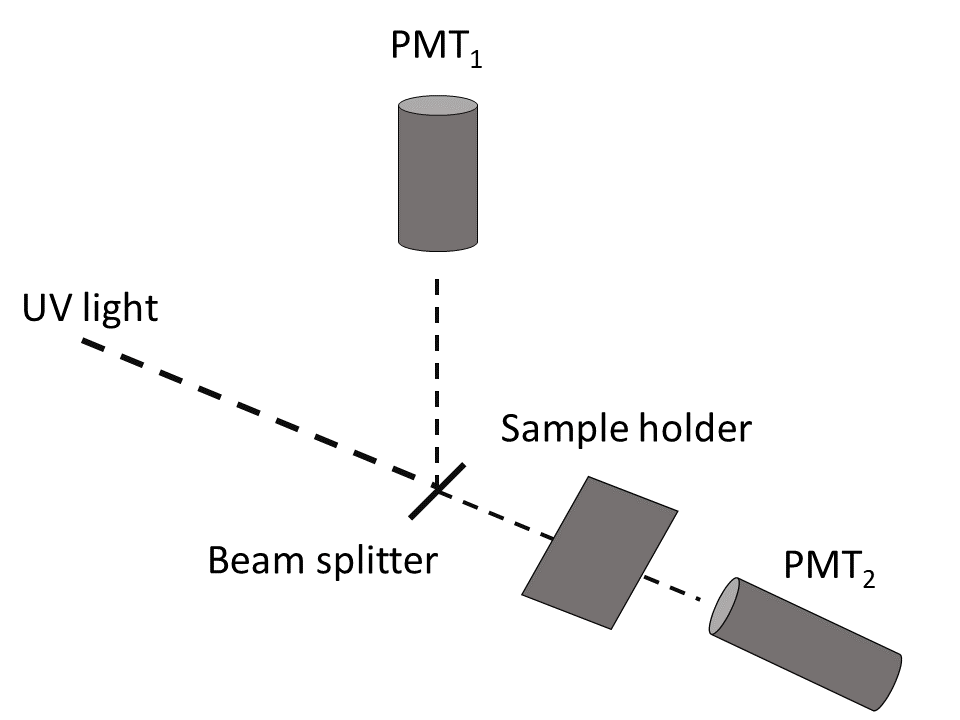}
    \caption{}
    \label{ASSET_step2}
  \end{subfigure}
  \caption{Schematic diagram of the two steps of QE measurement}
  \label{ASSETqetwosteps}
\end{figure}

\begin{figure}[ht]
  \centering
  \begin{subfigure}[b]{0.49\textwidth}
    \includegraphics[width=0.9\textwidth]{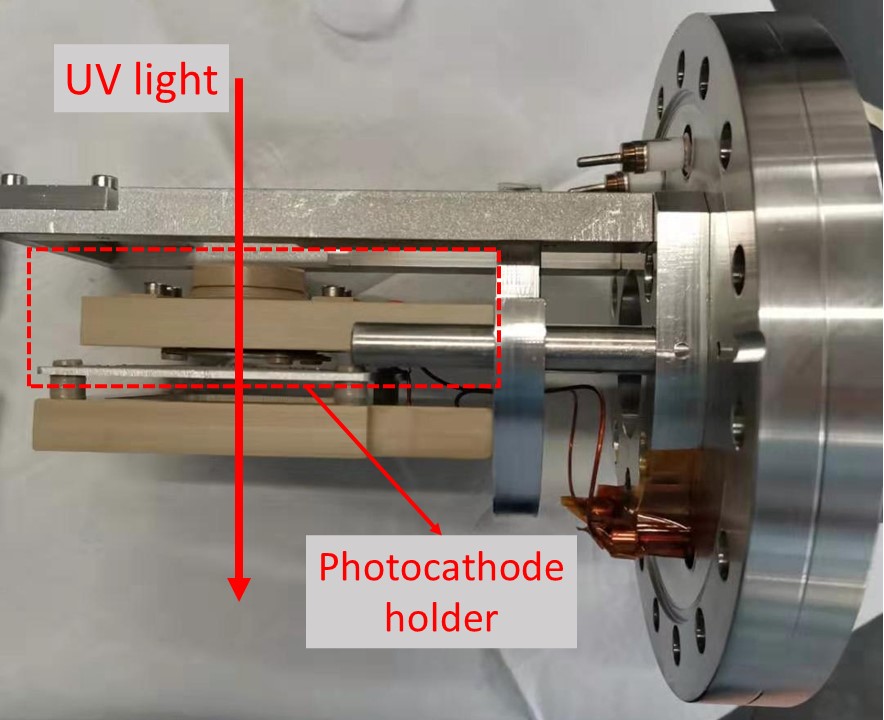}
    \caption{}
    \label{holder1}
  \end{subfigure}
  \hfill 
  \begin{subfigure}[b]{0.49\textwidth}
    \includegraphics[width=0.9\textwidth]{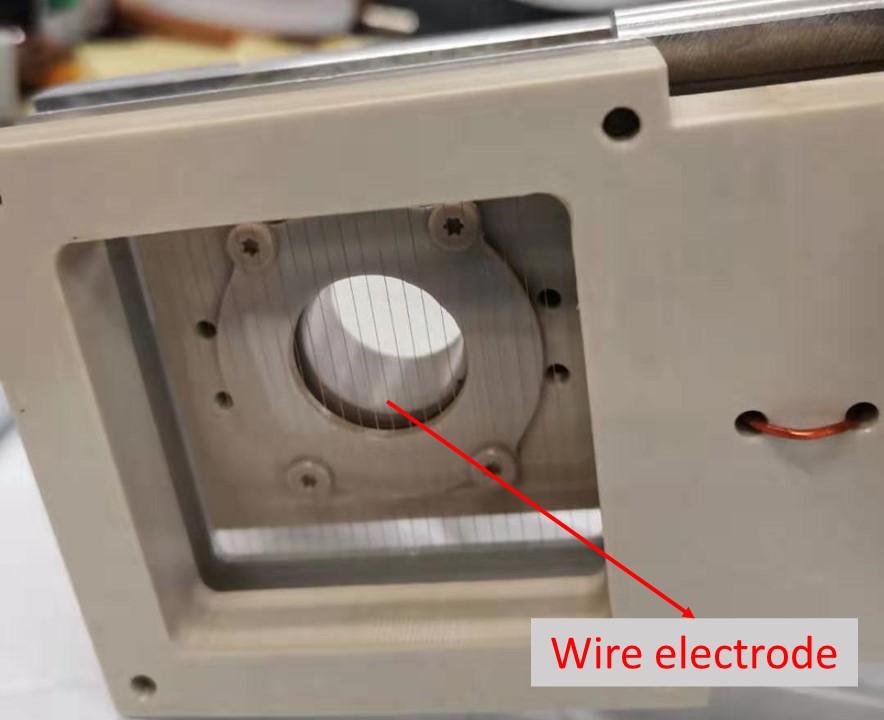}
    \caption{}
    \label{holder2}
  \end{subfigure}
  \caption{Pictures of the photocathode holder (a) and wire electrode (b) of the ASSET setup}
  \label{ASSETholder}
\end{figure}

Figure~\ref{ASSETqetwosteps} illustrates the two steps involved in the QE measurement using ASSET. As a first step, two PMTs are used to measure the intensity of two beams without photocathode in the light path at a given wavelength. Picoammeters (Keithley model 6482) are used to measure the current from the two PMTs, current $I_{ref0}$ from PMT$_1$ and $I_{abs0}$ from PMT$_2$ are recorded. Then, in the second step, a holder containing the photocathode and wire electrode, as shown in figure~\ref{ASSETholder}, is placed in the light path in front of the PMT$_2$ within the measurement chamber. A suitable voltage is applied to the wire electrode to establish an electric field between the wire electrode and the photocathode for PEs emission and collection. During this step, a picoammeter connected to the DLC film measures the photocurrent ($I_{pe}$) generated by the PEs emissions when the UV photons pass through the MgF$_2$ crystal and irradiate on the DLC film. The number of photoelectrons ($N_e$) emitted per second is calculated by dividing $I_{pe}$ by the charge of electron. At the same time, the current ($I_{ref1}$) generated from PMT$_1$ is recorded. The QE of the DLC photocathode can be obtained using the following formula:
\begin{equation}
QE = \frac{N_e}{N_p} = \frac{\frac{I_{pe}}{e}}{A\times I_{ref1} \times \frac{I_{abs0}}{I_{ref0}}}
	\label{qefunc}
\end{equation}
Where $N_p$ is the number of photons per second, $e$ is the charge of electron and $A$ is the response coefficient of PMT$_1$ to UV photons. In this case, the QE results contain the effects of MgF$_2$ crystal.\par

\subsection{Results}
\begin{figure}[ht]
\centering
\includegraphics[width=.6\textwidth]{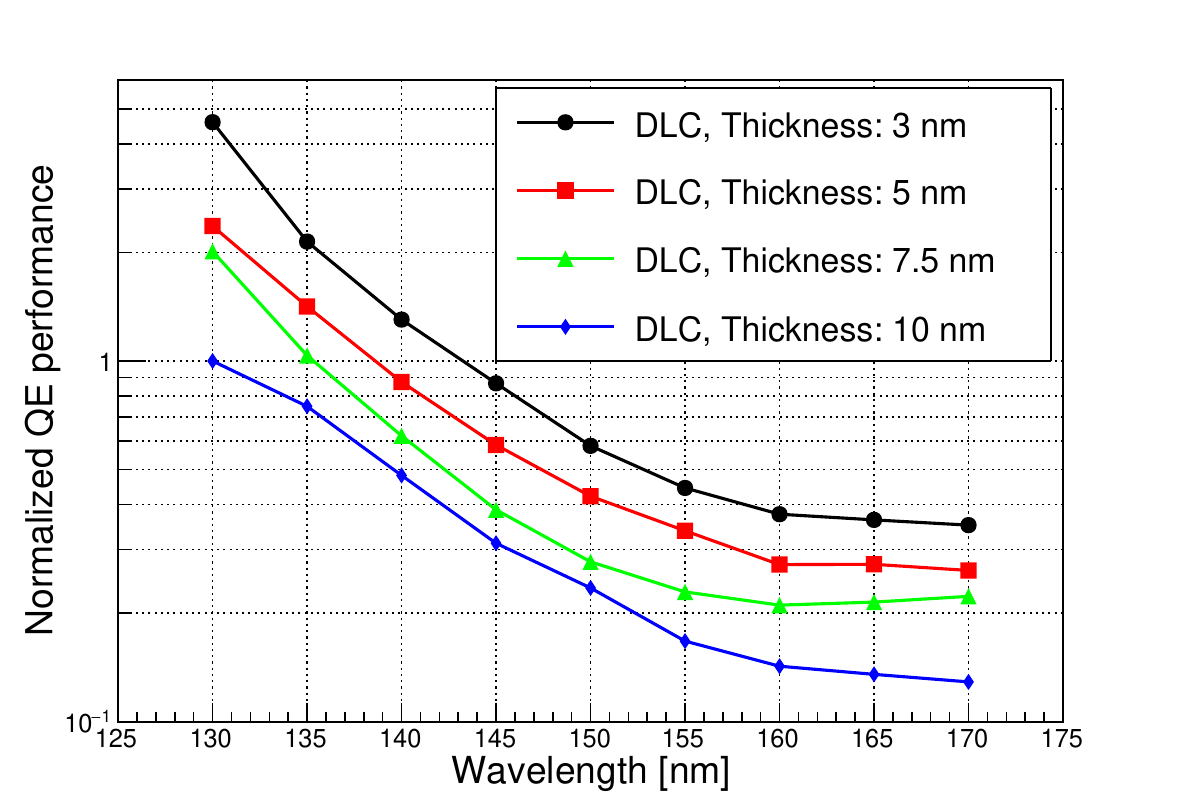}
\caption{Normalized QE performance of different DLC photocathodes}
\label{ASSET_DLC_NorQE}
\end{figure}

In this measurement, DLC photocathodes with different thicknesses were measured and the relative QE were compared. The QE results are shown in figure~\ref{ASSET_DLC_NorQE}, which corresponding to 3 \si{nm}, 5 \si{nm}, 7.5 \si{nm} and 10 \si{nm}, respectively. All of the results are normalized by using the QE value for the DLC photocathode of 10 \si{nm} that measured at the wavelength of 130 \si{nm}. The results indicate that the optimized thickness of the DLC photocathode is 3 \si{nm} and the QE value is the highest under 130 \si{nm} UV photons irradiation. Besides, attempts were made to produce and measure a thinner layer (about 1 \si{nm}) to further enhance the performance. However, both the production and QE measurements were not reliable, and the thinner layer was found lacked sufficient resistance for storage and operation.\par

\section{Aging test}
\label{Aging test}
Since the voltage of pre-amplification gap in the PICOSEC MM is high, the ratio of ions feedback to the position of photocathode is non-negligible. To verify the resistance of the DLC photocathode to ion bombardment, an aging test with a high intensity laser was carried out. A single channel PICOSEC MM prototype was developed, in which the photocathode was installed for the aging test.

\subsection{Detector prototype}
\begin{figure}[ht]
  \centering
  \begin{subfigure}[b]{0.565\textwidth}
    \includegraphics[width=\textwidth]{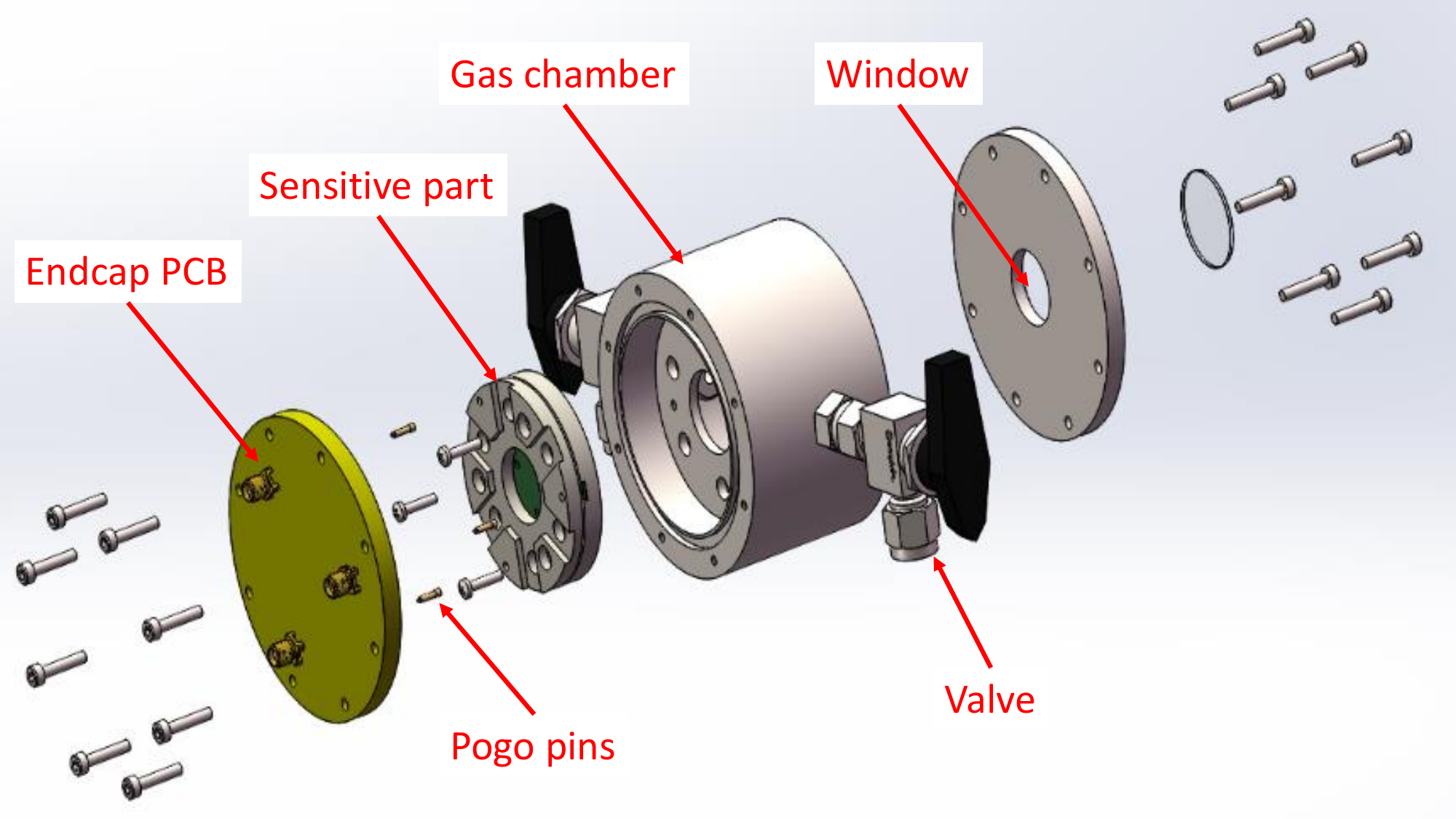}
    \caption{}
    \label{prototyoePICOSEC2}
  \end{subfigure}
  \hfill 
   \begin{subfigure}[b]{0.415\textwidth}
    \includegraphics[width=1\textwidth]{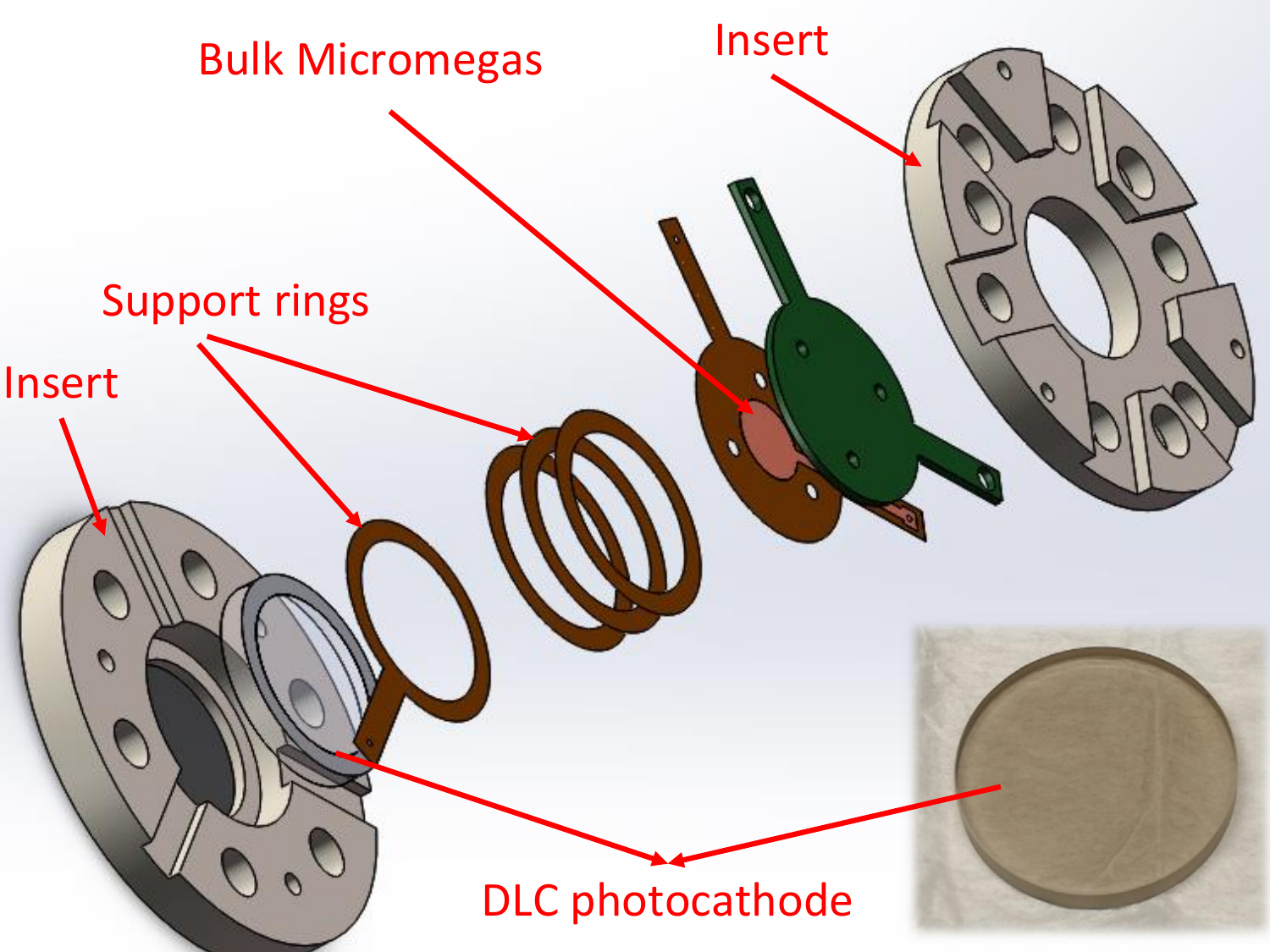}
    \caption{}
    \label{prototyoePICOSEC1}
  \end{subfigure}
  \caption{Exploded view of the single channel PICOSEC MM prototype (a) and the sensitive part of detector(b)}
  \label{prototypefigure}
\end{figure}

Figure~\ref{prototyoePICOSEC2} shows an exploded view of the single channel PICOSEC MM prototype. It consists of the endcap printed circuit board (PCB), the sensitive part,  the gas chamber and the quartz window, as well as screws and gas accessories. The sensitive part is fixed inside the gas chamber and its three electrode of photocathode, mesh and anode are connected to the endcap PCB through three pogo pins, respectively. The main components of the sensitive part are bulk MM sensor, support rings, photocathode and inserts, as shown in figure~\ref{prototyoePICOSEC1}. The bulk MM with a typical amplification gap of 128 \si{\micro\meter} is employed for electron multiplication and signal readout\cite{GIOMATARIS2006405}. Its effective area is 1 \si{cm} in diameter. The support rings are used to form a 200 \si{\micro\meter} pre-amplification gap between the photocathode and the mesh.

\subsection{Setup}
\begin{figure}[ht]
  \centering
  \begin{subfigure}[b]{0.49\textwidth}
    \includegraphics[width=1.\textwidth]{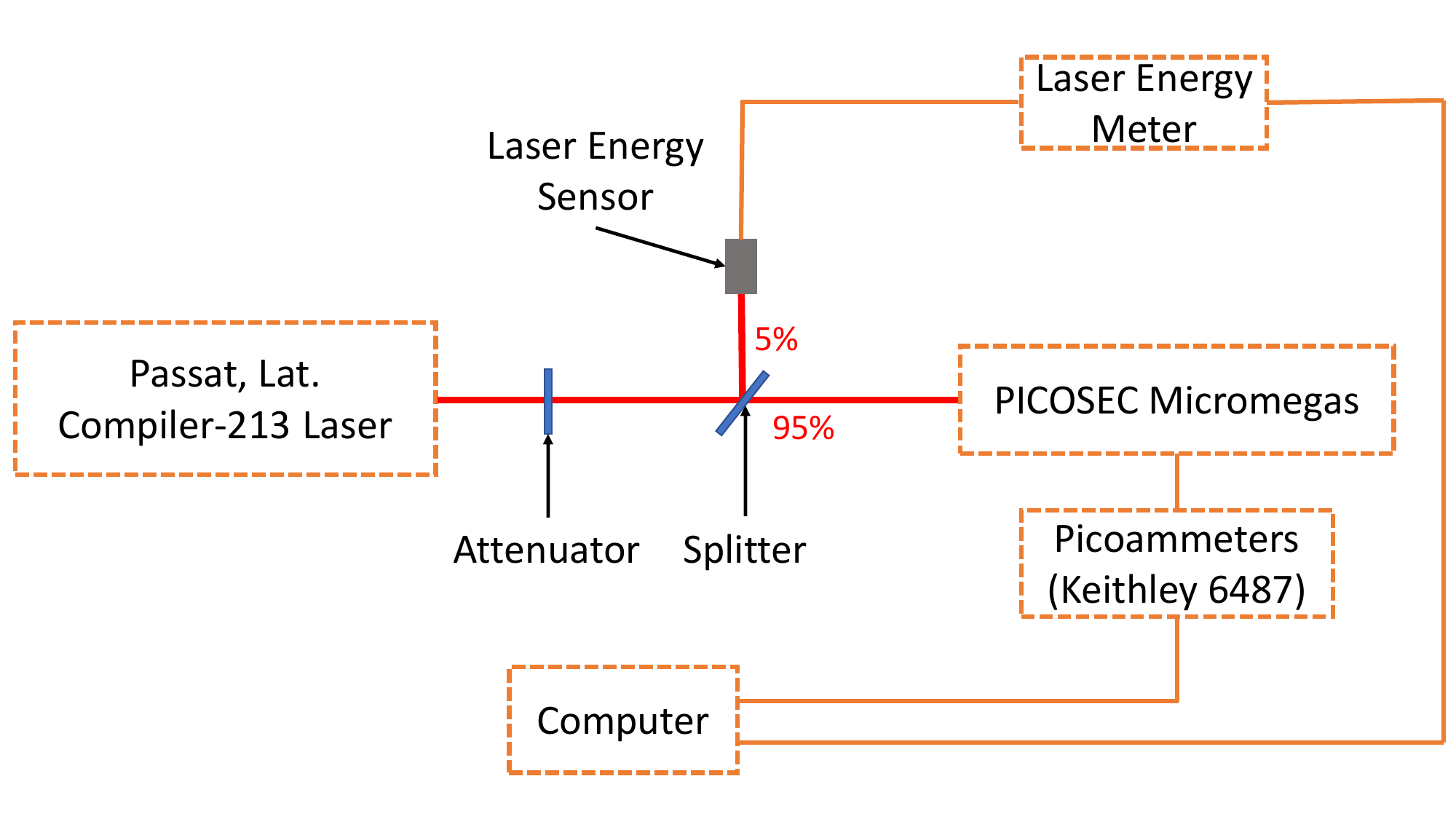}
    \caption{}
    \label{aging_schematic1}
  \end{subfigure}
  \hfill 
  \begin{subfigure}[b]{0.49\textwidth}
    \includegraphics[width=\textwidth]{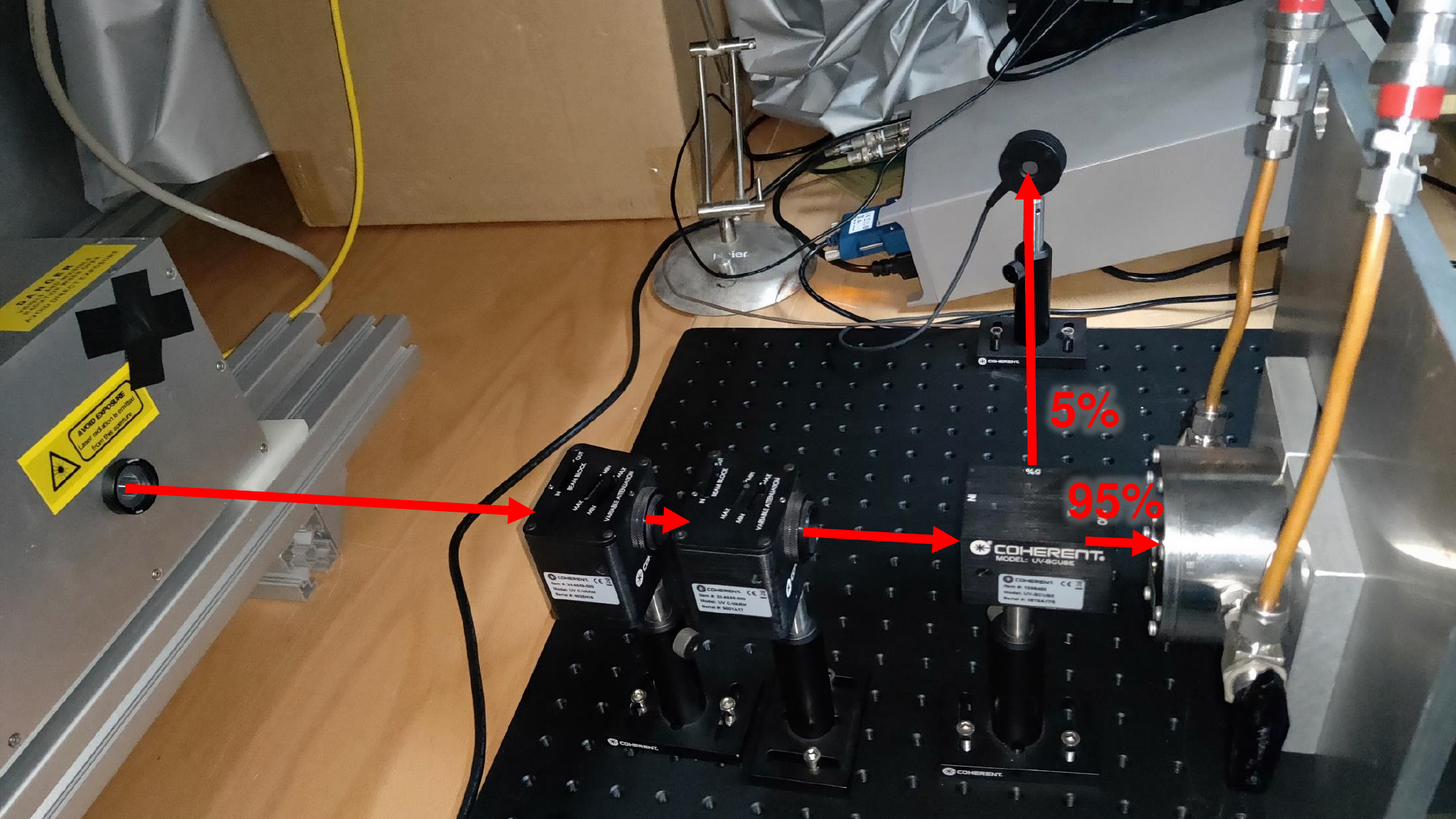}
    \caption{}
    \label{agingpicture1}
  \end{subfigure}
  \caption{Schematic diagram (a) and picture (b) of aging test setup}
  \label{agingpicture}
\end{figure}

The test setup is shown in figure~\ref{agingpicture}. A pulsed laser (Passat Lat. Compiler-213) was used to provide laser with a wavelength of 213 \si{nm}. The laser beam is split into two beams. One beam line is recorded by a laser energy meter (COHERENT PM10) which is used to monitor the stability of the laser intensity. The other beam line is sent to the prototype. Positive voltage is applied to the anode and mesh electrode, while the photocathode is maintained at 0 \si{V} and connected to a picoammeter (Keithley model 6487). The voltage of the pre-amplification gap and the amplification gap are set to 420 \si{V} and 300 \si{V}, respectively. The prototype maintains a constant gain by keeping the voltage and other parameters fixed.\par 

In the test, when the prototype is exposed to the laser, the photocathode promptly responds by emitting primary PEs, a direct consequence of the photoelectric effect. These PEs are then further ionized and multiplied in the gas gap, while the ions generated during the ionization process feedback to the position of photocathode and bombard the photocathode film. With the continuous irradiation of laser, the current of feedback ions $I(t)$ from the photocathode and the laser intensity $P(t)$ were simultaneously recorded by the picoammeter and the laser energy meter. Since the gas gain remains unchanged, the normalized current $I(t)/P(t)$ can be used as an indicator of the photocathode's QE performance. Additionally, the number of charges accumulated on the photocathode can be obtained by integrating the current $I(t)$. The visible degeneration area of the CsI photocathode after aging test was investigated under a microscope, and an irradiated area of approximately 0.06 cm$^2$ was obtained. This area differs from the one indicated in the manual, which is a circle with a diameter of 2.5 \si{mm}, due to the distortion caused by the laser passing through the optics and the window of the prototype.\par

\subsection{Results}
\begin{figure}[ht]
\centering
\includegraphics[width=.7\textwidth]{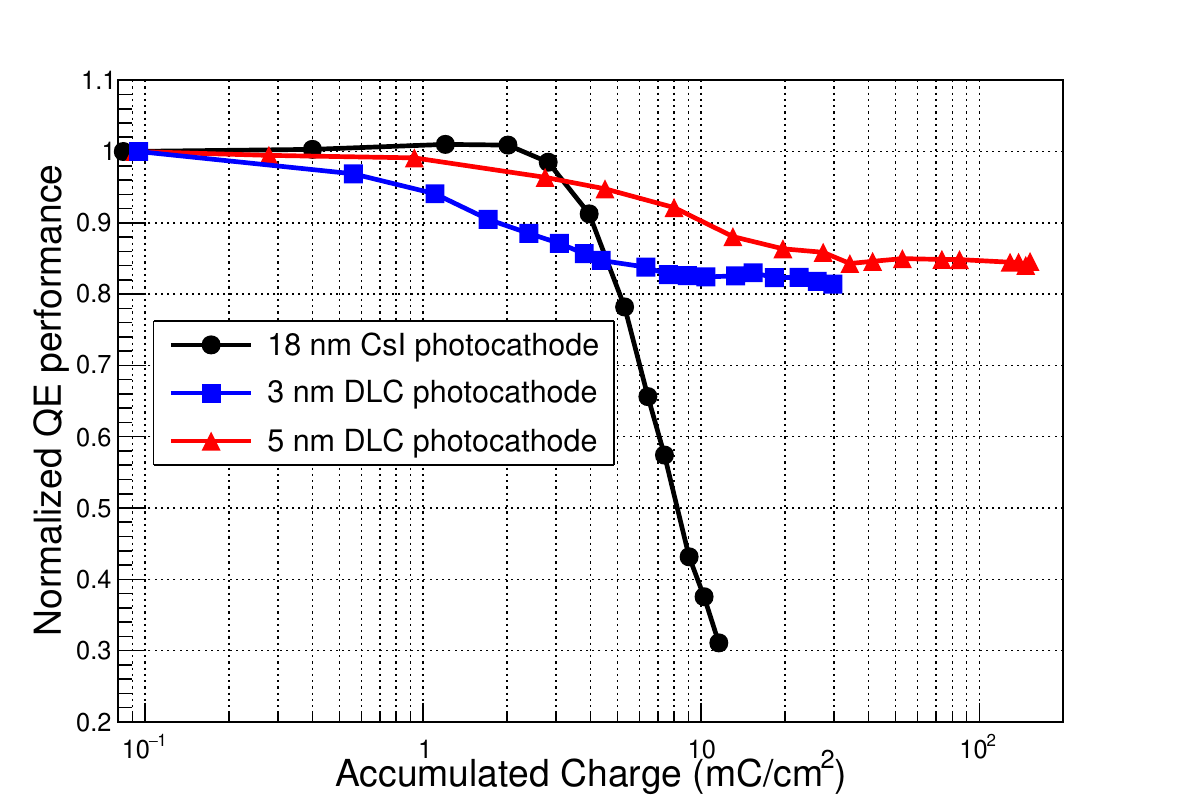}
\caption{Dependence of the normalized QE performance on the accumulated charges for different photocathodes}
\label{agingresults}
\end{figure}

The test results are shown in figure~\ref{agingresults}, where the x-axis represents the accumulated charge on the photocathode, and the y-axis shows the normalized QE performance of the photocathode. The three curves in figure~\ref{agingresults} correspond to one CsI photocathode and two DLC photocathodes, with thicknesses of 18 \si{nm} (black curve), 3 \si{nm} (blue curve) and 5 \si{nm} (red curve), respectively. For each curve, the first data point is normalized to 1. As can be seen from the plot, the performance of the CsI photocathode declines rapidly to around 30\% after being bombarded by ions of around 10 \si{mC/cm^2}. In contrast, the DLC photocathodes maintain stable QE performance after a slight deterioration. When the accumulated ions on the DLC photocathodes reach tens of \si{mC/cm^2}, or even exceed 100 \si{mC/cm^2}, their performance degradation is less than 20\%. Furthermore, the damaged region of the CsI photocathode after aging can be clearly seen under microscopic examination, which corresponds to the laser irradiation area, but it is very difficult to find marks on the DLC photocathodes when they are bombarded with even more feedback ions. These results indicate that the PICOSEC MM detector with DLC photocathodes can operate longer with stable performance compared to one with a CsI photocathode.\par

\section{Beam test}
\label{Beam test}
Besides the QE measurement and aging test, we have performed tests on the PICOSEC MM prototype equipped with DLC photocathodes in muon beams. These beam tests allowed us to evaluate the performance of DLC photocathodes in a near-application environment.\par

\subsection{Setup}
\begin{figure}[ht]
  \centering
  \begin{subfigure}[b]{0.9\textwidth}
    \includegraphics[width=1.\textwidth]{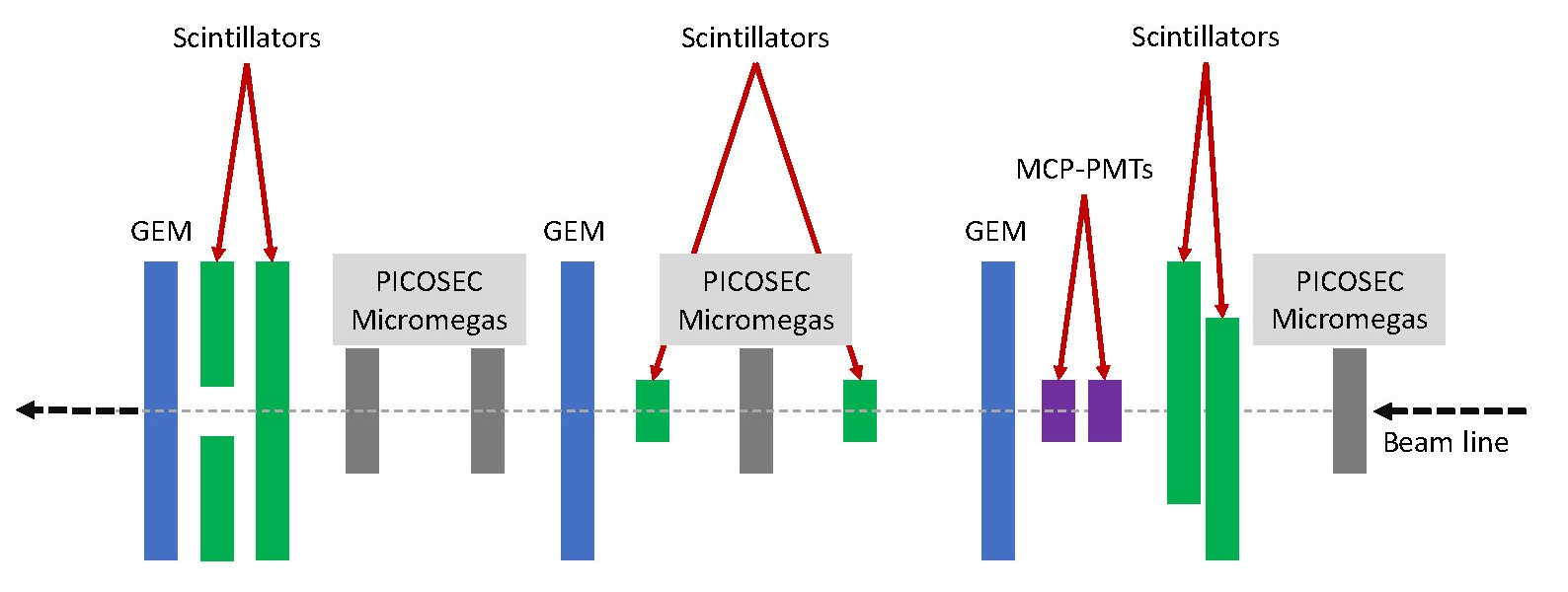}
    \caption{}
    \label{beamtelescope1}
  \end{subfigure}
  \hfill 
  \begin{subfigure}[b]{0.8\textwidth}
    \includegraphics[width=\textwidth]{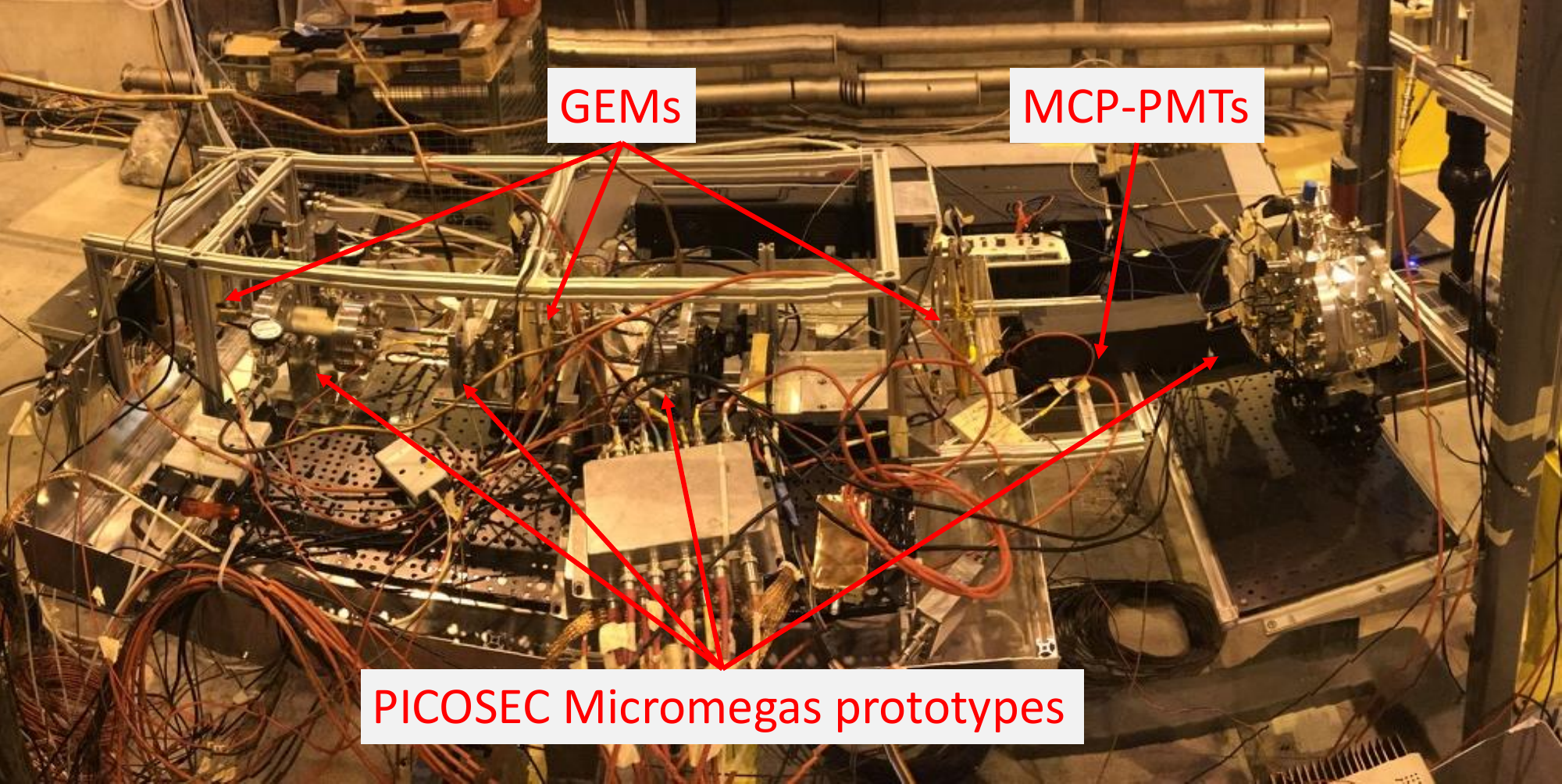}
    \caption{}
    \label{beamtelescope2}
  \end{subfigure}
  \caption{Schematic (a) and picture (b) of the beam test setup}
  \label{beamtelescope}
\end{figure}

Beam tests for the PICOSEC MM prototype were carried out at the CERN SPS H4 secondary line. 150 \si{GeV/c} muon beams were used to measure the performance of the PICOSEC MM prototype equipped with different DLC photocathodes. The beam test setup is shown in figure~\ref{beamtelescope}. It consists of a trigger system, a GEM tracker, time reference detectors and several positions for the PICOSEC MM prototypes. Different shaped scintillators are used to select the particle beam in a specific region and trigger the data acquisition system. Three triple-GEMs are used to provide precise tracking information of muons, and two multiple micro-channel plate photomultiplier tubes (MCP-PMT, model Hamamatsu R3809U-50) are employed as the time reference detectors which providing a time resolution of better than 6 \si{ps}\cite{sohl2019spatial}. The active area of the MCP-PMT is 11 \si{mm} in diameter, which is aligned with the anode electrode of the PICOSEC MM prototype, as well as the triggered region. In the data acquisition system, the signals of both the PICOSEC MM prototype and the MCP-PMs are recorded with an oscilloscope (20 \si{Gsps} and 4 \si{GHz} bandwidth). More details of the beam setup is shown in \cite{sohl2020development}.\par

In the beam test, the number of PEs extracted from the photocathode per muon incident (N$_{pe}$/$\mu$) was tested. The yield of PEs is closely related to the QE of the photocathode, which is the crucial performance of the photocathode. This number is estimated from the response of the muon beam and a single PEs calibration using a UV lamp at the same voltage. In addition, the time resolution of the prototype with different DLC photocathodes was tested to obtain the optimized time performance. The time information of the prototype and the MCP-PMT was obtained by fitting the leading edges with a sigmoid function and performing 20\% constant fraction discrimination (CFD). The time difference between the MCP-PMT and the prototype is defined as the “signal arrival time” (SAT). The time resolution ($\sigma$) of the prototype, which contains the contribution of the MCP-PMT and other effects, can be acquired by fitting the distribution of the SAT with a Gaussian function. This methods are described detailed in \cite{bortfeldt2018picosec, sohl2020development}. Both the results of yield of the DLC photocathodes and the time resolution of the prototype are detailed in the following sections.\par

\subsection{PEs yield}
The amplitude distributions of the prototype for single PEs and muon beam are fitted using a Polya function\cite{schindler2012microscopic,zerguerras2015understanding,sohl2020development}. The mean amplitude of single PEs and muons are denoted as $Q_e$ and $Q_{mu}$, respectively. The PEs yield is simply estimated by dividing $Q_{mu}$ by $Q_e$. The PEs yield results are shown in Table~\ref{table1}. When the thickness of DLC film is 3 \si{nm}, the optimal performance is obtained, for which the N$_{pe}$/$\mu$ equals to 3.7. By comparing all the photocathodes, it can be seen that as the thickness of the DLC film increases from 3 \si{nm} to 10 \si{nm}, the PEs yield decreases, which is consistent with the results of QE measurement. Additionally, under the same conditions, an 18 \si{nm} CsI photocathode was also tested, while the N$_{pe}$/$\mu$ was approximately 10\cite{bortfeldt2018picosec}.\par

\begin{table}
\centering
\caption{PEs yield results of different DLC photocathodes}
\centerline{
\begin{tabular}{|c|c|c|}
 \hline 
Thickness of DLC film (nm) &  N$_{pe}$/$\mu$  & Detection efficiency\\ 
\hline
3 & 3.7 & 97\%  \\
\hline
5 & 3.4 & 94\% \\
\hline
7.5 & 2.2 & 70\%\\
\hline
10 & 1.7 & 68\%\\
\hline
\end{tabular}}
\label{table1}
\end{table}

The more PEs extracted from the photocathode, the higher detection efficiency will be obtained. As the thickness of the DLC film increases from 3 \si{nm} to 10 \si{nm}, the detection efficiency of the prototype for charged particles decreases from 97\% to 68\%, while the detection efficiency approaches 100\% when using CsI photocathode. These results indicate that using a 3 \si{nm} DLC photocathode, the PICOSEC MM prototype can maintain sufficient detection efficiency for charged particles.\par

\subsection{Time resolution}
\begin{figure}[ht]
  \centering
  \begin{subfigure}[b]{0.49\textwidth}
    \includegraphics[width=1.\textwidth]{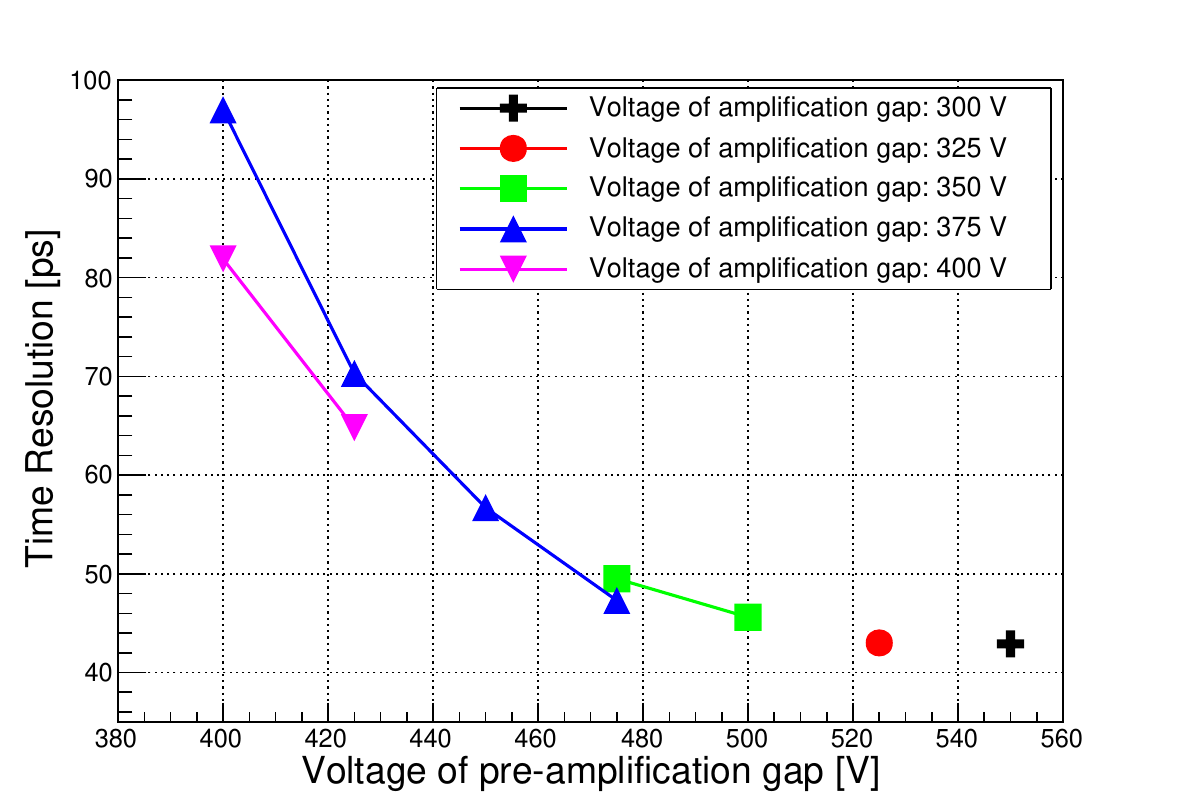}
    \caption{}
    \label{Beam_3DLCnoaging_time}
  \end{subfigure}
  \hfill 
  \begin{subfigure}[b]{0.49\textwidth}
    \includegraphics[width=\textwidth]{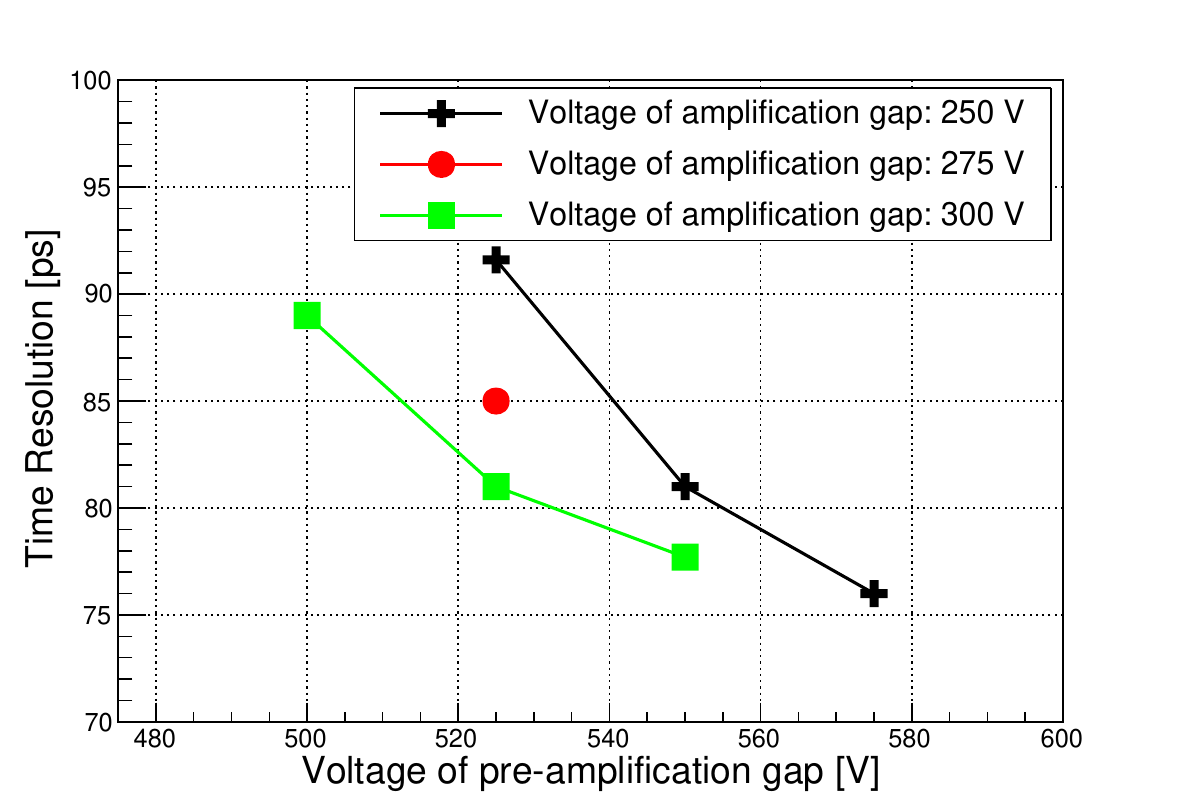}
    \caption{}
    \label{Beam_10DLC_time}
  \end{subfigure}
  \caption{Time resolution results of the PICOSEC MM prototype equipped with a 3 \si{nm} and 10 \si{nm} DLC photocathode}
  \label{DLCtimeresolution}
\end{figure}

The time resolution of the prototype equipped with different types of DLC photocathode was tested. Of these, the 3 \si{nm} DLC photocathode exhibits the best performance, followed by the 5 \si{nm}, 7.5 \si{nm}  and 10 \si{nm} photocathodes in descending order of performance, which is consistent with the results of PEs yield and QE. Figure~\ref{Beam_3DLCnoaging_time} and \ref{Beam_10DLC_time} show the time resolution of the prototype equipped with 3 \si{nm} and 10 \si{nm} DLC photocathodes, respectively. It can be clearly seen that the time resolution improves with higher pre-amplification voltage, which has the same tendency when using CsI photocathode. With the 3 \si{nm} DLC photocathode, the time resolution reaches up to approximately 42 \si{ps} at the voltage of 550 \si{V} of pre-amplification gap and 300 \si{V} of amplification gap. With the thickness of the DLC film increases, the time resolution deteriorated. With the 10 \si{nm} DLC photocathode, the best time resolution is only about 76 \si{ps}, even higher voltage on pre-amplification gap was applied.\par

\section{Conclusion}
The DLC photocathode for the PICOSEC MM detector is introduced, and it has shown promising results with its suitable sensitivity and stability. Its performance is optimized through the manufacturing processes, QE measurements and beam tests. In the QE measurements, samples with different thickness were studied, and the results indicate that the optimized thickness of DLC photocathode is approximately 3 \si{nm}. Besides, the aging effect of ion bombardment was also tested. Both tested DLC photocathodes maintain stable QE performance after a slight decrease, which is less than 20\%, after being bombarded with feedback ions of around 100 \si{mC/cm^2}. As a comparison, the CsI photocathode has shown rapid damage when only 10 \si{mC/cm^2} is accumulated on the CsI film. Additionally, muon beams was used to test the PICOSEC MM prototype equipped with DLC photocathodes to verify its performance in practical application environment. A time resolution of around 42 \si{ps} with a detection efficiency of 97\% for 150 \si{GeV/c} muons were obtained when the prototype equipped with a 3 \si{nm} DLC photocathode. In conclusion, the good properties of no hydrolysis, chemical stability and robustness to ion bombardment of the DLC photocathode demonstrate its great application potential as the photocathode for the PICOSEC MM detector.\par

\acknowledgments

We acknowledge the support of the Program of National Natural Science Foundation of China (grant number 11935014, 12125505, 12075238); the CERN EP R\&D Strategic Programme on Technologies for Future Experiments; the RD51 collaboration, in the framework of RD51 common projects; the Cross-Disciplinary Program on Instrumentation and Detection of CEA, the French Alternative Energies and Atomic Energy Commission; the PHENIICS Doctoral School Program of Université Paris-Saclay, France; the COFUND-FP-CERN-2014 program (grant number 665779); the Fundaç\~ao para a Ciência e a Tecnologia (FCT), Portugal (CERN/FIS-PAR/0005/2021); the Enhanced Eurotalents program (PCOFUND-GA-2013-600382); the US CMS program under DOE contract No. DE-AC02-07CH11359. The authors wish to thank the Hefei Comprehensive National Science Center for their support.

\bibliographystyle{JHEP}
\bibliography{biblib.bib}
\end{document}